# Assessing Validity of Static Analysis Warnings using Ensemble Learning

Anshul Tanwar, Hariharan Manikandan, Krishna Sundaresan, Prasanna Ganesan,
Sathish Kumar Chandrasekaran, Sriram Ravi

*Cisco Systems*

## Abstract

Static Analysis (SA) tools are used to identify potential weaknesses in code and fix them in advance, while the code is being developed. In legacy codebases with high complexity, these rules-based static analysis tools generally report a lot of false warnings along with the actual ones. Though the SA tools uncover many hidden bugs, they are lost in the volume of fake warnings reported.

The developers expend large hours of time and effort in identifying the true warnings. Other than impacting the developer productivity, true bugs are also missed out due to this challenge. To address this problem, we propose a Machine Learning (ML)-based learning process that uses source codes, historic commit data, and classifier-ensembles to prioritize the True warnings from the given list of warnings. This tool is integrated into the development workflow to filter out the false warnings and prioritize actual bugs. We evaluated our approach on the networking C codes, from a large data pool of static analysis warnings reported by the tools. Time-to-time these warnings are addressed by the developers, labelling them as authentic bugs or fake alerts. The ML model is trained with full supervision over the code features. Our results confirm that applying deep learning over the traditional static analysis reports is an assuring approach for drastically reducing the false positive rates.

## 1. Introduction

Static Analysis (SA) tools are used across organizations to detect code flaws early in the cycle. Various open-source ([5-7]) and commercial ([8]) SA tools identify programming errors. The major limiting factor of these tools is the percentage of false positive warnings that get reported. Since the SA tools are rules and heuristic-based, in enterprise software with higher complexity false positive rates increase drastically.

Several prior research works apply ML to reduce these false warnings. But these approaches do not directly analyze source code. They consider the meta attributes of code like code complexity, size of code, frequency of bugs, etc as the feature-set to describe a SA warning. These features when used to classify the given code warning, might not be very

- *Anshul Tanwar - PRINCIPAL ENGINEER E-mail: atanwar@cisco.com.*
- *Hariharan Manikandan - SOFTWARE ENGINEER E-mail: hmanikan@cisco.com*
- *Krishna Sundaresan - VP ENGINEERING E-mail: ksundar@cisco.com*
- *Prasanna Ganesan - DIRECTOR ENGINEERING E-mail: prasgane@cisco.com*
- *Sathish Kumar Chandrasekaran - TECHINCAL LEADER E-mail: sathicha@cisco.com*
- *Sriram Ravi – MANAGER ENGINEERING E-mail: srravi@cisco.com*

accurate, since they are not direct data representations of the source code that contains the warnings.

The static analysis warnings can be further categorized based on the Common Weakness Enumeration (CWE) they fall under. We hypothesize that the false positive rates can vary greatly depending on the domain of the code and has a high correlation to the associated CWE enumeration. Based on the nature of the code, the false positive percentages reported for a given CWE category can vary for different software. Our goal is to consider these CWE types in conjunction with the code features and to build ML models.

In the proposed approach, the first step is to learn vector representations of source codes using an attention-based model [1]. Second, we build dedicated binary classifiers for each category of CWE. For each CWE dataset, the true and false SA warning codes are vectorized and classified into 0 or 1. Post-training, the ML models are added to the production workflow. In this phase, the trained ML offers developer productivity gains by prioritizing SA warnings that have a high chance of being a bug. It cuts downtime, via automatic first-hand assessment of warnings, contingent on knowledge of historic data.

Based on the code commit history, a dataset of SA warning codes, their fixes, and the falsely flagged bug-free codes were mined. From the results of the evaluation, it was observed that the models achieved 97% recall and 92% accuracy for the top occurring CWE categories. Additionally, the predictions generated on a test dataset of open SA warnings were further validated with the domain experts. The real-time F1-score on test samples of the top ten most frequent CWEs was recorded at 82.72%. The classifiers were also hyperparameter tuned with grid search. The best performing models with optimal parameters were deployed for real-time assessment. The details of the training and fine-tuning will be shared in later sections.

The rest of the manuscript will be organized in the following sections. In the related work section, we focus on few existing methods that address a similar problem. In the data preparation section, we discuss the data sources and feature extraction schemes. The AI modelling section covers the different classification models trained and tested for the datasets. In the last section, the results achieved and the deployment strategies are discussed.

## 2. Related Works

A distributed representation for codes was proposed in [1]. This code2vec approach uses an attention-based neural network to learn representative vectors for programs. Similar to word vectors, source code can be vectorized and consumed for downstream activities like classification.

The SARD Juliet test suite is a benchmarked dataset of synthetic test cases used to validate the efficacy of various static analysis tools [9]. Koc et. al employed the combination of error reports from multiple tools as features to train ML models on the Juliet test cases [2]. The classifier is fit to a labelled dataset of code flaws from Juliet, which are first analyzed by the SA tools. The SA analyzed is matched with ground truth labelling and the alerts are identified as fake or real.

Flynn et. al built a SA alert classifier over open-source Juliet programs [3]. First, a program slice from the line of error is extracted and program-specific tokens are replaced with generic symbols. This code data normalization is done so that the program is generalizable across different datasets. A Long Short Term Memory (LSTM) network is used as a classifier. This study is done mainly on Java code slices for SQL injection flaws. In [4], Zachary et. al analyzed the SA warnings messages to create a pattern code for identifying false positive warnings.

In this work, we propose ML for C-language SA warning prioritization using code embeddings learned from the source code. To the best of our knowledge, this ML for SA validity assessment from code vector representations is the first of its kind.

## 3. Data Preparation and Feature Extraction

To effectively classify the different types of code flaws we need a vast amount of balanced dataset. The data should also have sufficient samples across the warning categories for the classifiers to work across all types of flaws.

**Table 1.** Listing of dataset counts under each CWE category of SA warnings from historic data.

| S. No | Common Weakness Enumeration | Number of true warning samples | Number of fixed programs | Number of fake alert samples | Total dataset size for training and validation | Test dataset size |
|---|---|---|---|---|---|---|
| 1 | CWE-476 | 26567 | 19393 | 2246 | 48206 | 40499 |
| 2 | CWE-404 | 4233 | 2836 | 681 | 7750 | 2358 |
| 3 | CWE-561 | 1484 | 742 | 2057 | 4283 | 11919 |
| 4 | CWE-252 | 1365 | 1228 | 535 | 3128 | 2644 |
| 5 | CWE-119 | 1178 | 415 | 1474 | 3067 | 2971 |
| 6 | CWE-457 | 1222 | 794 | 886 | 2902 | 2395 |
| 7 | CWE-394 | 900 | 585 | 53 | 1538 | 290 |
| 8 | CWE-590 | 416 | 400 | 608 | 1424 | 1313 |
| 9 | CWE-125 | 522 | 208 | 287 | 1017 | 2139 |
| 10 | CWE-667 | 264 | 200 | 175 | 639 | 2064 |

We collected the historic user commits for the purpose of identifying and prioritizing true positive SA warnings. Table 1 enumerates the counts of data under each CWE bucket. The true positive warnings were the bugs reported by the tool and fixed by the developers. The false positive data comes from the developer dismissing the warning as a fake alert. These sets are labelled as 1 and 0 respectively. In addition to that, the fixed codes submitted by the developer after resolving a real SA warning are also included as part of data creation for training. These samples present no trace of security warnings, therefore considered as false positive. These synthetic false positives (codes without any SA warnings) were used to augment the dataset counts.

The two main objectives considered during the data collection are (i) recall for true positive class should be the topmost priority, followed by (ii) reduction in false positives reported. The goal of this activity is to enable the model to learn the nuances of the fix and draw a finer line of separation with respect to true warnings.

This categorical data presented in Table 1, was used to build an array of binary classifiers for each CWE type. The overall dataset has 60K+ labelled rows spread across 10 CWE categories. The curated dataset was split in an 80:20 ratio to form training and validation data. Besides, a set of existent SA warnings with no history of user action is taken as a testing set. These are open warnings that are not fixed by development teams. These are unseen real-world data, on which the model's real performance can be robustly evaluated. The predictions generated on this set are verified by the security domain experts and the efficacy of the models is established.

With the source code data and labels, the dataset is framed. We convert source code to AST and convert the AST to code embeddings. This code vectorization step is derived from an attention neural network, the Code2Vec model, that was pre-trained for classifying functionally similar codes [1]. These learned Code2vec model parameters are leveraged for creating code vectors from AST. These code vector representations are used as features to train AI to check if the code presents any evidence for the SA warning. The Code2vec feature dimensionality, which is a configurable hyperparameter, was empirically chosen to be equal to 384 in all experiments.

The code embeddings were created for function-level codes. Since the code vectors are learned from the AST, their latent dimensions are known to precisely express the semantic structure of codes. These code vectors are used as the predominant features for the classification task.

## 4. Ensemble Modelling

The real and fake SA warnings are bucketed into broader CWE categories. Bootstrap sampling with replacement is applied to the original dataset to create bootstrapped datasets. A heterogeneous array of binary classifiers are learned over each bootstrapped data partition. The predictions generated by the models are aggregated to determine the final result. The final classification is derived as a majority voting of classification probabilities from these base learners. The three base ML learners in the heterogeneous bagging ensemble are the Extreme Gradient Boost machine, 5-layer Neural Network, and Random Forests. These classifiers learn to specialize in identifying true positive alerts reported by the SA tools for that CWE set. The ensemble models are optimized with a search for ideal hyperparameters. Figure 1 illustrates the steps involved in setting up the ML for training these classifiers. In the figure, for each component learner, the attributes searched are mentioned against the classifier algorithm. The majority class voting scheme for combining the predictions yielded better results than a single classification from any one of the three models.

The ensemble was instantiated and trained for each individual CWE dataset comprising of real and fake warnings. The models were learned on an 80:20 training-validation dataset split. The trained models were run over a test set of open SA warnings. These open issues are freshly reported by the SA tool and are not a part of the data fed into ensemble learning. We used the active learning approach where the domain experts are asked to evaluate the model prediction on this new unlabelled warnings set.

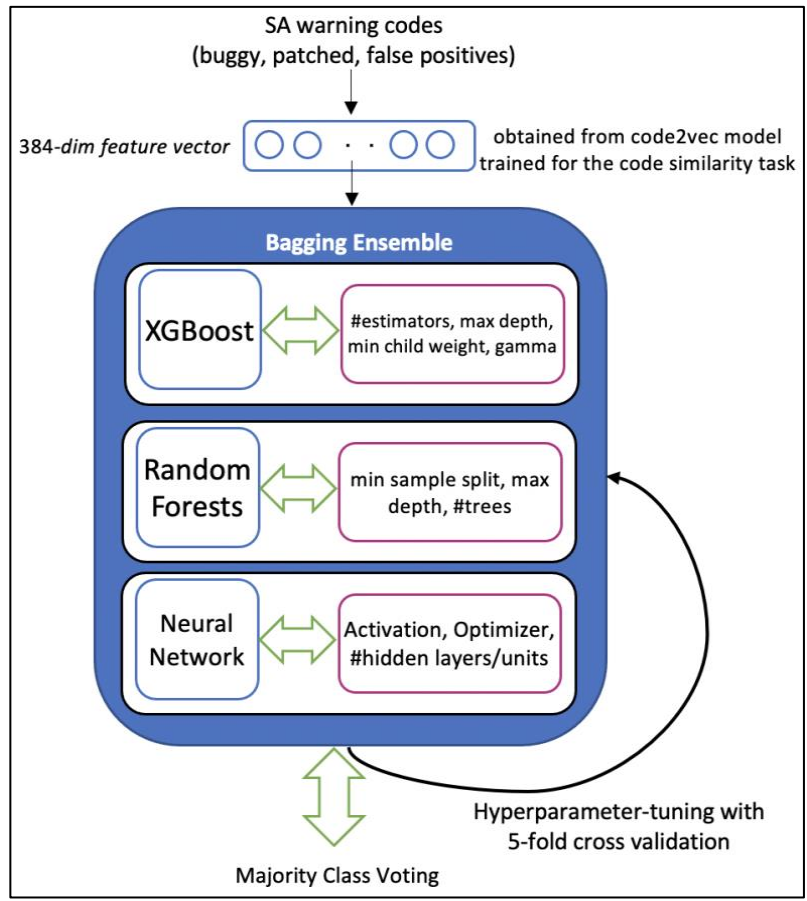

Fig. 1: Architectural Diagram of the proposed classifier ensemble.

### 4.1. Model Fine-tuning

The ten CWEs listed in table 1 are a well-defined collection of historic codes for training the ensemble model. Further, to arrive at the best configuration of the model, the ML trained on each CWE dataset was hyperparameter tuned using Grid Search. The hold-out validation dataset was used to evaluate the choice of values. The hyperparameter settings that maximize the performance on this set were considered ideal for real-world prediction. Listed in table 2 are the search parameters of the classifiers which were optimized to achieve high ensemble classification efficacy on the validation set. All possible combinations of the values in the specified ranges were tried to get the best results.

Table 2: Range of hyperparameter values searched for fine-tuning.

| S. No. | Base learner | Hyperparameter | Grid Search values |
|---|---|---|---|
| 1 | Extreme Gradient Boosting | Maximum depth of learner decision tree | {3, 5, 7, 9} |
|   |   | Minimum child weight for the split | {1, 3, 5} |

|   |   | L2 regularization parameter | {0.1, 0.2, 0.3, 0.4, 0.5} |
|---|---|---|---|
| 2 | Neural Network | Learning rate decay | {0.95, 0.5, 0.1} |
|   |   | Optimizer | {Adam, SGD, AdaDelta} |
|   |   | Number of hidden layers | {2,3,4,5} |
|   |   | Number of units in each layer | {128, 256, 512} |
| 3 | Random Forests | Minimum number of samples to split at a tree node | {2, 4, 6} |
|   |   | Max depth of the tree | {5, 10, 15} |
|   |   | Number of estimators | {10, 100, 500} |

The hyperparameter tuned models were validated by security experts. This human-in-the-loop approach helped us in improving the classifiers that lacked appropriate data and also improve false positive predictability.

**4.2. Deployment and Usage**

There are three places where the proposed models can create real value:

- SA warnings fix in Legacy Code: For a very long time, there are some warnings left open in certain pockets of code due to high false positive reporting from the SA tools. The model predictions are leverage to create a priority order for addressing the legacy SA warnings clean-up. The high-priority warnings as predicted by the ML can be potential security bugs, which may pose serious threats. They can be resolved amidst the possibly false positive ones.

- Developer Workflow: We have integrated our ML augmented Static analysis system into the developer commit workflow. Whenever a developer writes a new block of code and runs static analysis tools on the code, our model would create the code vectors and run them through the bagging ensemble to provide real-time ML score feedback to the developer.

- Improving developer judgment of the SA warnings: Since SA tools emit too many false positives, the developer's mindset has come to dismiss most of the warnings without paying proper attention to the descriptions. The ML can be used to place a check on this form of user behavior, wherein the true warnings incorrectly dismissed as false positives can be highlighted by the ML to revisit the developer's judgment.

**4.3. Active Learning**

An active learning loop for refining the ML has been established in the production workflow. As developers work their way through the list of SA warnings, they would be assisted with the insights from the learned ML. This saves their time from being wasted in inspecting false positives. The developer actions are tracked and the model is actively fine-tuned for the CWEs where real-world performance was found lacking. This continuous learning process works both ways, i.e. firstly the model learns from the direct and indirect developer feedbacks on the warnings they are working on, secondly, the developers are well-informed to re-evaluate the false positives they have marked when the model prediction is otherwise.

## 5. Results Analysis

Table 3 provides the empirical results of model training and validation. The metric scores were registered on the validation set. Since the test dataset is unlabelled, the model's performance was studied on the 20% unseen validation split. These ML ensemble classifiers built on CWE datasets were robustly evaluated on these ML metrics: Precision, Recall, Accuracy and F1-score. The average precision and recall for these 10 CWEs were 82.725% and 83.765%. This shows that the ensemble models were more than 80% effective in detecting true SA warnings. The recall scores are a direct reflection of the model's ability to capture real vulnerabilities. CWE-404, CWE-457, CWE-119, CWE-590 have displayed the highest recall for correctly identifying most of the warnings at a very low chance of emitting false positives. These weakness enumerations deal with improper resource shutdown, uninitialized variable usage, buffer errors, and improper free of memory. The area under the ROC describes the extent of separability of true positive warnings from fake SA alerts. Across the top CWE categories, the area under ROC is over 79.23% on average. The F1 scores are an overall reflection of the model's degree of fit to the dataset. With an average F1 score of 82.72%, it is revealed that the ML has captured the real and falsely reported SA warning patterns well.

**Table 3.** Performance Analysis of the classifier ensemble trained for different CWE datasets.

| S. No | Common Weakness Enumeration | Accuracy | Precision | Recall | F1-score | Area under ROC |
|---|---|---|---|---|---|---|
| 1 | CWE-476 | 86.77 | 81.40 | 89.95 | 85.46 | 88.86 |
| 2 | CWE-404 | 80.94 | 85.17 | 93.25 | 89.02 | 84.27 |
| 3 | CWE-561 | 76.93 | 80.94 | 86.15 | 83.46 | 80.35 |
| 4 | CWE-252 | 85.84 | 88.71 | 79.17 | 83.66 | 79.64 |
| 5 | CWE-457 | 92.56 | 94.24 | 94.43 | 94.33 | 94.86 |
| 6 | CWE-119 | 80.07 | 83.35 | 84.18 | 83.76 | 77.77 |
| 7 | CWE-394 | 74.39 | 75.42 | 73.63 | 74.51 | 72.98 |
| 8 | CWE-125 | 75.84 | 78.55 | 81.93 | 80.20 | 73.03 |
| 9 | CWE-590 | 87.31 | 87.81 | 91.61 | 89.66 | 84.71 |
| 10 | CWE-667 | 62.43 | 63.03 | 63.35 | 63.19 | 56.48 |

One of the unique design strategies that resulted in such high scores was the learning rate decay. Especially for the neural network model, the learning rate was decayed by a factor of 10% over a patience range of 5 epochs. This ensured excellent model convergence.

The prioritization scores emitted by the models were categorized into high, medium, and low bands by fitting the scores into a normal distribution. The threshold for the high range is set at a confidence of 95% and for the medium priority range, it was 66%. The development teams were encouraged to start addressing the SA warnings in the higher band, followed by medium and low priority ones.

## 5. Conclusion

This work presented a new technique to train accurate and robust ML models from AST-based code vectors for Static Analysis warnings validation. The proposed work minimizes the development team's time and effort in inspections SA warnings, by forming a priority order upon these warnings list. The code2vec algorithm was used to obtain deep AST-based feature representations for code. These vectors were a result of training an AST path attention neural network on the code's AST structure. The obtained vectors are classified using a bagging ensemble consisting of an extreme gradient boosting machine, random forests, and neural network.

The proposed approach also presents a way to retrain the model iteratively based on user feedback. Upon user validation, selective CWE classifiers are hyperparameter tuned and deployed back into production. The ML models yielded remarkable results in SA warnings assessment for 10 CWE categories with an average F1-score of 82.725%. As future work, the models can be extended to include the set of CWEs published as top security threats across the software industry.

**Acknowledgments**

The authors thank Cisco Systems for allowing us to carry out this research work and also, we would like to thank the domain experts and developers who helped us manually validate the initial results of the model. It helped us fine tune the models for production deployment.